\documentclass[superscriptaddress,nofootinbib,a4paper,11pt]{article}
\usepackage{jheppub}
\usepackage{graphicx}
\usepackage{subcaption}
\usepackage{float}
\usepackage{amsmath}
\usepackage{amssymb}
\usepackage[utf8]{inputenc}
\usepackage{hyperref}
\usepackage{color}
\usepackage{slashed}
\usepackage{multirow}
\usepackage{enumitem}
\usepackage{lineno}
\allowdisplaybreaks

\begin{document}

\title{\boldmath 
Does the Weinberg angle allow a local hidden-variable description for the leptonic decays of an entangled $ZZ$ pair?
}

\author[a,*]{Junle Pei \note[*]{Corresponding author.}}

\affiliation[a]{Institute of Physics, Henan Academy of Sciences, Zhengzhou 450046, P. R. China}

\emailAdd{peijunle@hnas.ac.cn}

\abstract{
Quantum entanglement in di-boson systems offers a useful testing ground for exploring the boundary between quantum-mechanical correlations and classical descriptions based on local hidden variables. In this work, I study the spin-polarization state of a $Z_1Z_2$ pair produced from the decay of a spin-0 particle and investigate whether the angular correlations predicted by quantum field theory (QFT) in the leptonic decays $Z_1(\to e^-_1 e^+_1)Z_2(\to e^-_2 e^+_2)$ can be reproduced by a local hidden-variable theory (LHVT) under angular-momentum conservation. By matching the LHVT angular distribution to the QFT prediction coefficient by coefficient, I derive the conditions under which an LHVT construction exists. 
For the case $w\neq 0$, I show that, apart from trivial product-state configurations, an LHVT construction exists only for a unique entangled state, corresponding to $a_1=a_3=-a_2=1/\sqrt{3}$ and $b_2=b_3=0$, together with restricted windows of the weak-mixing angle $\theta_W$. For $w=0$, I derive a necessary and sufficient criterion for the existence of an LHVT construction in terms of a closed set of algebraic and positivity conditions. As an application, I consider the phenomenologically relevant interaction $sZ^\mu Z_\mu$ and show that an LHVT construction exists at threshold $m_s=2m_Z$, whereas it does not exist for $m_s>2m_Z$.
}

\maketitle

\section{Introduction} \label{sec:int} 

Quantum entanglement is one of the most distinctive features of quantum theory \cite{PhysRev.47.777,PhysicsPhysiqueFizika.1.195,PhysRevLett.28.938,PhysRevLett.47.460,PhysRevLett.49.1804,Bouwmeester_1997,PhysRevLett.81.3563,PhysRevLett.80.1121,Riebe:2004jpa,Barrett:2004bxh}, and its manifestation in high-energy processes has attracted increasing attention in recent years \cite{BESIII:2018cnd,Fabbrichesi:2021npl,Afik:2020onf,Han:2023fci,Cheng:2023qmz,Subba:2024mnl,Subba:2024aut,ATLAS2024,CMS:2024pts,Han:2025ewp,vonKuk:2025kbv,Pei:2025non,Pei:2025yvr,Lin:2025eci,Wu:2025dds,BESIII:2025vsr,Cheng:2025cuv}. Compared with the traditional settings of quantum information and atomic physics, collider and decay processes offer a qualitatively different environment in which entanglement can be generated, propagated, and eventually probed through the angular correlations of final-state particles. In particular, the polarization correlations of unstable vector bosons provide a natural laboratory for studying entanglement in relativistic quantum field theory (QFT) \cite{Subba:2024aut,Goncalves:2025mvl,Goncalves:2025xer,Goncalves:2026njf}.

Among such systems, the $Z_1Z_2$ pair produced in the decay of a spin-0 particle is especially simple and theoretically clean. Angular-momentum conservation strongly constrains the allowed polarization structure of the di-boson state, while the subsequent leptonic decays of the two $Z$ bosons make the spin information experimentally accessible through fully differential angular distributions. This makes the process $Z_1(\to e_1^-e_1^+)Z_2(\to e_2^-e_2^+)$ a useful framework for investigating the relation between quantum correlations and possible classical descriptions.

Inspired by the discussions in Refs.~\cite{ABEL1992304,Li:2024luk,Bechtle:2025ugc,Abel:2025skj,Low:2025aqq,Pei:2026wfu,Pei:2026khg},
the central question is whether the angular correlations in the leptonic decays $Z_1(\to e^-_1 e^+_1)Z_2(\to e^-_2 e^+_2)$, as predicted by QFT, can be reproduced by a local hidden-variable theory (LHVT). In the present work, I address this question for $Z_1Z_2$ states originating from the decay of a spin-0 parent particle. Our analysis is based on two ingredients. First, I write the QFT prediction for the joint angular distribution of the leptonic final states in a spherical-harmonic basis, so that the full spin information of the $Z_1Z_2$ state is encoded in a set of angular coefficients. Second, I construct a general LHVT under the assumption of angular-momentum conservation, in which the hidden variables are identified with the spin directions of the two intermediate $Z$ bosons in their rest frames. The hidden-variable distribution and the decay response function are then expanded in spherical harmonics and Legendre polynomials, respectively, allowing for a direct coefficient-by-coefficient comparison with the QFT result.

This strategy leads to a sharp characterization of the parameter region in which an LHVT can exist. For $Z_1Z_2$ states produced from spin-0 decay, the QFT polarization state is restricted to the form $|Z_1Z_2\rangle = a_1|-1\rangle_{Z_1}\otimes|-1\rangle_{Z_2}
+ a_2 e^{i b_2}|0\rangle_{Z_1}\otimes|0\rangle_{Z_2}
+ a_3 e^{i b_3}|1\rangle_{Z_1}\otimes|1\rangle_{Z_2}$,
subject to normalization. I determine, within this parameter space, when the LHVT angular distribution can exactly reproduce the QFT prediction. The problem naturally separates into the two cases $w\neq 0$ and $w=0$, where $w$ characterizes the parity-violating structure of the $Z\ell\ell$ coupling. As I show below, the former case is highly restrictive and admits only a unique nontrivial solution with restricted intervals of the weak-mixing angle $\theta_W$, whereas the latter allows a broader class of solutions that can nevertheless be characterized by a complete set of necessary and sufficient conditions.

Our analysis is also motivated by phenomenology. In particular, for the commonly encountered interaction structure $sZ^\mu Z_\mu$, the polarization state of the produced $Z_1Z_2$ system is entangled and fixed by the masses of the parent scalar and the $Z$ boson. This makes it possible to determine directly whether an LHVT construction exists in physically relevant regimes. I find that the threshold configuration $m_s=2m_Z$ admits such a construction, while for $m_s>2m_Z$ the QFT angular correlations cannot be reproduced by any LHVT consistent with all criteria required.

The remainder of this paper is organized as follows. In Sec.~\ref{sec:2}, I present the QFT description of the processes of $Z_1(\to e^-_1 e^+_1)Z_2(\to e^-_2 e^+_2)$ and derive the angular distribution of the final-state leptons. In Sec.~\ref{sec:3}, I formulate the LHVT under angular-momentum conservation and determine the conditions under which it can reproduce the QFT result for the leptonic decays of $Z_1Z_2$ produced from spin-0 decay. I conclude in Sec.~\ref{sec:con}.

\section{Leptonic angular distribution in 
$Z_1Z_2$ decays: QFT description and entanglement criterion} \label{sec:2} 

I consider the process $Z_1(\to e^-_1e^+_1)Z_2(\to e^-_2e^+_2)$. For a pure $Z_1Z_2$ state, the spin-polarization state can be written as
\begin{align}
|Z_1Z_2\rangle=\sum_{k=-1}^1\sum_{j=-1}^1\alpha_{k,j}|k\rangle_{Z_1}\otimes|j\rangle_{Z_2}~, \label{z12}
\end{align}
where $k$ and $j$ are the spin-projection quantum numbers of $Z_1$ and $Z_2$, respectively, defined along their own momentum directions. The coefficients $\alpha_{k,j}$ satisfy
\begin{align}
\sum_{k=-1}^1\sum_{j=-1}^1 |\alpha_{k,j}|^2=1~. \label{akj}
\end{align}

Without loss of generality, I perform the analysis in the center-of-mass (c.m.) frame of the $Z_1Z_2$ system throughout this work. The momentum direction of $Z_1$, denoted by $\hat{e}_{Z_1}$, is chosen as the polar axis of the spherical coordinate system, namely the $\theta=0$ direction. An arbitrary unit vector $\hat{e}_{\perp}$ orthogonal to $\hat{e}_{Z_1}$ is taken to define the $\phi=0$ direction and $\phi$ increases according to the right-hand rule around $\hat{e}_{Z_1}$.
For the decay process $Z_i \to e^-_i+ e^+_i~(i=1,2)$, I denote by $\hat{e}_i$ the momentum direction of $e_i^-$ in the rest frame of $Z_i$, and parametrize it as
\begin{align}
    \hat{e}_i=(\sin\theta_i\cos\phi_i,\sin\theta_i\sin\phi_i,\cos\theta_i)~.
\end{align}
The leptonic configuration is therefore fully specified by the four angles $(\theta_1,\phi_1,\theta_2,\phi_2)$.

I denote the joint angular distribution of $\hat{e}_1$ and $\hat{e}_2$ by $\mathcal{W}(\theta_1,\phi_1,\theta_2,\phi_2)$. According to QFT, for the pure $Z_1Z_2$ state described by Eq.~(\ref{z12}), $\mathcal{W}(\theta_1,\phi_1,\theta_2,\phi_2)$ is given by
\begin{align}
&\mathcal{W}(\theta_1,\phi_1,\theta_2,\phi_2)\nonumber\\
=&\frac{1}{N}\sum_{k=-1}^1\sum_{j=-1}^1\sum_{m=-1}^1\sum_{n=-1}^1\sum_{\lambda_1=-1}^1\sum_{\lambda_2=-1}^1\alpha_{k,j}\alpha^*_{m,n}\mathcal{M}_1(k,\lambda_1)\mathcal{M}_1^*(m,\lambda_1)
\mathcal{M}_2(j,\lambda_2)\mathcal{M}_2^*(n,\lambda_2)~.
\end{align}
Here, $N$ is a normalization factor chosen such that
\begin{align}
    \int_{-1}^1 d\cos\theta_1 \int_{0}^{2\pi}d\phi_1
\int_{-1}^1 d\cos\theta_2 \int_{0}^{2\pi}d\phi_2~\mathcal{W}(\theta_1,\phi_1,\theta_2,\phi_2)=1~.
\end{align}
The helicity amplitudes are given by
\begin{align}
    &\mathcal{M}_1(k,\lambda_1)
    =e^{i k\phi_1}d^1_{\lambda_1,k}(\theta_1)h_{\lambda_1}~,\quad \mathcal{M}_2(j,\lambda_2)
    =e^{-i j\phi_2}d^1_{\lambda_2,-j}(\theta_2)h_{\lambda_2}~,
\end{align}
where $\lambda_i=\lambda_{e^-_i}-\lambda_{e^+_i}=0,\pm 1$, and $\lambda_{e^{-}_i}, \lambda_{e^{+}_i}=\pm \frac{1}{2}$ are the helicities of $e^{-}_i$ and $e^{+}_i$, respectively, defined along their own momentum directions in the $Z_i$ rest frame. $d^1_{\lambda_i,k}(\theta_i)$ is the Wigner-d function.
$h_{\lambda_i}$ is given by\footnote{This result is obtained in the limit where the electron mass is neglected.}
\begin{align}
   & h_0=0~,\quad h_{-1}=1-2\sin^2\theta_W~,\quad h_1=2\sin^2\theta_W~,
\end{align}
where $\theta_W$ denotes the Weinberg angle.
By defining
\begin{align}
    w=\frac{h_{-1}^2-h_{1}^2}{h_{-1}^2+h_{1}^2}=\frac{-1+2\cos(2\theta_W)}{2-2\cos(2\theta_W)+\cos(4\theta_W)}~,
\end{align}
I can rewrite $\mathcal{W}(\theta_1,\phi_1,\theta_2,\phi_2)$ as
\begin{align}
\mathcal{W}(\theta_1,\phi_1,\theta_2,\phi_2)=&\frac{9}{64\pi^2}\sum_{k=-1}^1\sum_{j=-1}^1\sum_{m=-1}^1\sum_{n=-1}^1\sum_{\lambda_1=\pm 1}\sum_{\lambda_2=\pm 1}\alpha_{k,j}\alpha^*_{m,n}(1-\lambda_1 w)(1-\lambda_2 w)\nonumber\\
&\times e^{i(k-m)\phi_1} e^{i(-j+n)\phi_2}d^{1}_{\lambda_1,k}(\theta_1)d^{1}_{\lambda_1,m}(\theta_1)
d^{1}_{\lambda_2,-j}(\theta_2)d^{1}_{\lambda_2,-n}(\theta_2)~.
\end{align}

For an arbitrary observable $\mathcal{O}(\theta_1,\phi_1,\theta_2,\phi_2)$, its expectation value can be written as
\begin{align}
&\langle\mathcal{O}(\theta_1,\phi_1,\theta_2,\phi_2)\rangle=\sum_{k=-1}^1\sum_{j=-1}^1\sum_{m=-1}^1\sum_{n=-1}^1\mathcal{O}_{k,j,m,n}\alpha_{k,j}\alpha^*_{m,n}~,
\end{align}
with
\begin{align}
\mathcal{O}_{k,j,m,n}=&\int_{-1}^1 d\cos\theta_1 \int_{0}^{2\pi}d\phi_1
\int_{-1}^1 d\cos\theta_2 \int_{0}^{2\pi}d\phi_2~\mathcal{O}(\theta_1,\phi_1,\theta_2,\phi_2)e^{i(k-m)\phi_1} e^{i(-j+n)\phi_2}\times \nonumber\\
&
\sum_{\lambda_1=\pm 1}\sum_{\lambda_2=\pm 1} (1-\lambda_1 w)(1-\lambda_2 w) d^{1}_{\lambda_1,k}(\theta_1)d^{1}_{\lambda_1,m}(\theta_1)
d^{1}_{\lambda_2,-j}(\theta_2)d^{1}_{\lambda_2,-n}(\theta_2)~.
\end{align}
This form cleanly separates the information associated with the production dynamics, encoded in $\alpha_{k,j}$, from the angular kernel associated with the observable itself. Once the coefficients $O_{k,j,m,n}$ are determined, the corresponding expectation values follow immediately for any given spin state of the $Z_1Z_2$ system.
For example,
\begin{align}
    \langle\cos(2\phi_1- 2\phi_2)\rangle=\frac{1}{8}\Delta~,\quad
\Delta=\alpha_{-1,-1}\alpha_{1,1}^*+\alpha_{1,1}\alpha_{-1,-1}^*~.
\end{align}
Under the normalization condition given in Eq.~(\ref{akj}), $\Delta\in[-1,1]$, including mixed states. By contrast, for unentangled $Z_1Z_2$, the allowed range is $\Delta\in[-\frac{1}{2},\frac{1}{2}]$, again valid for mixed states. Therefore,
\begin{align}
    8\langle\cos(2\phi_1- 2\phi_2)\rangle\in [-1,\frac{1}{2})\cup (\frac{1}{2},1]
\end{align}
provides a sufficient condition for quantum entanglement in $Z_1Z_2$. Similar criteria can be constructed from other observables.

\section{LHVT analysis of $Z_1 Z_2$ from spin-0 decay} \label{sec:3} 

\subsection{LHVT parameterization under angular‑momentum conservation}\label{sec:3-1}

When $Z_1Z_2$ originates from the decay of a spin-0 particle $s$, namely, $s\to Z_1+Z_2$, angular momentum conservation implies that, within a QFT description, the polarization state of the resulting pure $Z_1Z_2$ system must take the form
\begin{align}
& |Z_1Z_2\rangle=a_1|-1\rangle_{Z_1}\otimes|-1\rangle_{Z_2}+a_2 e^{i b_2}|0\rangle_{Z_1}\otimes|0\rangle_{Z_2}+a_3e^{i b_3}|1\rangle_{Z_1}\otimes|1\rangle_{Z_2}~. \label{sz1z2}
\end{align}
The pure $Z_1Z_2$ state in Eq.~(\ref{sz1z2}) is necessarily entangled as long as at least two of the $a_i~(i=1,2,3)$ are nonzero \cite{Pei:2025non}.
Without loss of generality, I choose
\begin{align}
& \quad a_1\in [0,1]~, \quad a_{2,3}\in [-1,1]~, \quad b_2\in (-\frac{\pi}{2},\frac{\pi}{2}]~, \quad b_3\in (-\frac{\pi}{2},\frac{\pi}{2}] ~. \label{cskj1}
\end{align}
The normalization condition requires
\begin{align}
& a_1^2+a_2^2+a_3^2=1~. \label{cskj2}
\end{align}
Eqs.~(\ref{cskj1}) and (\ref{cskj2}) define the full parameter space to be explored in the following discussion. 
 
I denote by $\mathcal{W}_s^\text{QFT}(\theta_1,\phi_1,\theta_2,\phi_2)$ the joint angular distribution of the final-state particles obtained in QFT for the $Z_1Z_2$ state in Eq.~(\ref{sz1z2}), and expand it in spherical harmonics as
\begin{align}
\mathcal{W}_s^\text{QFT}(\theta_1,\phi_1,\theta_2,\phi_2)=\sum_{l_1=0}^{\infty}\sum_{m_1=-l_1}^{l_1}\sum_{l_2=0}^{\infty}\sum_{m_2=-l_2}^{l_2}\frac{\mathcal{C}_{l_1,m_1,l_2,m_2}}{(2l_1+1)(2l_2+1)(-1)^{l_2}}Y^*_{l_1,m_1}(\hat{e}_1)Y^*_{l_2,m_2}(\hat{e}_2)~.
\end{align}
The values of $4\pi\mathcal{C}_{l_1,m_1,l_2,m_2}$ are listed in Table~\ref{tab:1}. For each coefficient shown there, the corresponding coefficient with interchanged indices satisfies
\begin{align}  \mathcal{C}_{l_2,m_2,l_1,m_1}=\mathcal{C}_{l_1,m_1,l_2,m_2}^*~.
\end{align}
All other coefficients $\mathcal{C}_{l_1,m_1,l_2,m_2}$ with $l_1\le l_2$ that are not listed in the table vanish. In particular, all $\mathcal{C}_{l_1,m_1,l_2,m_2}$ with $m_1+m_2\ne 0$ are zero.

\begin{table}[htbp]
\centering
\renewcommand{\arraystretch}{2}
\begin{tabular}{cccc}
\hline
$l_1,m_1$ & $l_2,m_2$ & $4\pi\mathcal{C}_{l_1,m_1,l_2,m_2}$  & $4\pi\mathcal{C}^\prime_{l_1,m_1,l_2,m_2}$ \\
\hline
$0,0$ & $0,0$ & $1$ &  $1$ \\
$0,0$ & $1,0$ & $\frac{3\sqrt{3}}{2}(a_1^2-a_3^2)w$ &  $2c_1 g_{1,0}\sqrt{\pi}$ \\
$0,0$ & $2,0$ & $\frac{\sqrt{5}}{2}(1-3a_2^2)$ & $2c_2 g_{2,0}\sqrt{\pi}$  \\
$1,-1$ & $1,1$ & $\frac{27}{4}a_2(a_1 e^{i b_2}+a_3 e^{-i(b_2-b_3)})w^2$ &  $\frac{1}{5}c_1^2(-5+2g_{2,0}\sqrt{5\pi})$ \\
$1,0$ & $1,0$ & $\frac{27}{4}(1-a_2^2)w^2$ &  $\frac{1}{5}c_1^2(5+4g_{2,0}\sqrt{5\pi})$ \\
$1,-1$ & $2,1$ & $\frac{9\sqrt{5}}{4}a_2(a_1 e^{ib_2}-a_3 e^{-i(b_2-b_3)})w$ & $c_1c_2(6g_{3,0}\sqrt{\frac{\pi}{35}}-2g_{1,0}\sqrt{\frac{3\pi}{5}})$ \\
$1,0$ & $2,0$ & $\frac{3\sqrt{15}}{4}(a_1^2-a_3^2)w$ & $\frac{2}{7}c_1c_2(14g_{1,0}+3\sqrt{21}g_{3,0})\sqrt{\frac{\pi}{5}}$ \\
$1,1$ & $2,-1$ & $\frac{9\sqrt{5}}{4}a_2(a_1 e^{-ib_2}-a_3 e^{i(b_2-b_3)})w$ & $c_1c_2(6g_{3,0}\sqrt{\frac{\pi}{35}}-2g_{1,0}\sqrt{\frac{3\pi}{5}})$ \\
$2,-2$ & $2,2$ & $\frac{15}{2}a_1 a_3e^{ib_3}$ & $\frac{1}{7}c_2^2(7+2g_{4,0}\sqrt{\pi}-4g_{2,0}\sqrt{5\pi})$ \\
$2,-1$ & $2,1$ & $\frac{15}{4}a_2(a_1 e^{ib_2}+a_3 e^{-i(b_2-b_3)})$ & $\frac{1}{7}c_2^2(-7+8g_{4,0}\sqrt{\pi}-2g_{2,0}\sqrt{5\pi})$ \\
$2,0$ & $2,0$ & $\frac{5}{4}(1+3a_2^2)$ & $\frac{1}{7}c_2^2(7+12g_{4,0}\sqrt{\pi}+4g_{2,0}\sqrt{5\pi})$ \\
\hline
\end{tabular}
\caption{$\mathcal{C}_{l_1,m_1,l_2,m_2}$ and $\mathcal{C}^\prime_{l_1,m_1,l_2,m_2}$.}
\label{tab:1}
\end{table}

I now turn to the corresponding local hidden-variable description. In this framework, the spin angular momentum of $Z_i~(i=1,2)$ in its rest frame is denoted by $\vec{S}_i$. For the $Z_1Z_2$ pair produced in the decay $s\to Z_1Z_2$, angular momentum conservation requires
\begin{align}
    & \vec{S}_{1}=-\vec{S}_{2}~.
\end{align}
Let $\hat{S}_i$ be the unit vector along $\vec{S}_i$. I parameterize it as
\begin{align}
    & \hat{S}=\hat{S}_{1}=-\hat{S}_{2}=(\sin x \cos y,\sin x \sin y,\cos x)~. \label{s1s2}
\end{align}
Here, $x\in[0,\pi]$ and $y\in[0,2\pi]$ are taken as the local hidden variables. Their distribution is described by a nonnegative normalized function $G(x,y)$ satisfying
\begin{align}
   & G(x,y)\ge 0~, \quad \int_{-1}^1 d\cos x\int_0^{2\pi}d y~ G(x,y)=1~. \label{gyhg}
\end{align}
Expanding $G(x,y)$ in spherical harmonics, one has
\begin{align}
    G(x,y)=\sum_{L=0}^{\infty}\sum_{M=-L}^{L}g_{L,M}Y_{L,M}(x,y)~. \label{Gxy}
\end{align}
The normalization condition in Eq.~(\ref{gyhg}) then implies
\begin{align}
    g_{0,0}=\frac{1}{\sqrt{4\pi}}~.
\end{align}

For a vector boson $Z_i$ at rest with spin pointing along $\hat{S}_i$, the angular distribution of the final-state particle $e_i^-$ in the decay $Z_i\to e_i^-e_i^+$ is described by a function $F(\hat{S}_i\cdot \hat{e}_i)$ satisfying
\begin{align}
   & F(\hat{S}_i\cdot \hat{e}_i)\ge 0~,\quad \int_{-1}^1 d\cos\theta_i\int_{0}^{2\pi} d\phi_i ~F(\hat{S}_i\cdot \hat{e}_i)=1~. \label{gyhf}
\end{align}
It is convenient to expand this function in Legendre polynomials as
\begin{align}
   & F(\hat{S}_i\cdot \hat{e}_i)=\frac{1}{4\pi}\sum_{l=0}^{\infty} c_l P_l(\hat{S}_i\cdot \hat{e}_i)~.\label{Fse}
\end{align}
The normalization condition in Eq.~(\ref{gyhf}) requires
\begin{align}
    c_0=1~.
\end{align}
In addition, the following identity holds:
\begin{align}
    P_l(\hat{S}_i\cdot \hat{e}_i)=\frac{4\pi}{2l+1}\sum_{m=-l}^l Y_{l,m}(\hat{S}_i)~Y^*_{l,m}(\hat{e}_i)~. \label{hdsse}
\end{align}

Using Eqs.~(\ref{s1s2}), (\ref{Gxy}), (\ref{Fse}), and (\ref{hdsse}), the joint angular distribution predicted by the local hidden-variable theory, denoted by $\mathcal{W}^\text{LHVT}_s(\theta_1,\phi_1,\theta_2,\phi_2)$, can be written as \cite{Pei:2026wfu}
\begin{align}
&\mathcal{W}^\text{LHVT}_s(\theta_1,\phi_1,\theta_2,\phi_2)=\int_{-1}^1d \cos x\int_0^{2\pi}d y~ G(x,y)~F(\hat{S}\cdot\hat{e}_1) ~F(-\hat{S}\cdot\hat{e}_2) \\
&=\sum_{l_1=0}^{\infty}\sum_{m_1=-l_1}^{l_1}\sum_{l_2=0}^{\infty}\sum_{m_2=-l_2}^{l_2}\frac{\mathcal{C}^\prime_{l_1,m_1,l_2,m_2}}{(2l_1+1)(2l_2+1)(-1)^{l_2}}Y_{l_1,m_1}^*(\hat{e}_{1})~Y_{l_2,m_2}^*(\hat{e}_{2})~.
\end{align}
Here,
\begin{align}
    \mathcal{C}^\prime_{l_1,m_1,l_2,m_2}=
    c_{l_1} c_{l_2} \left\langle Y_{l_1,m_1}(\hat{S})~Y_{l_2,m_2}(\hat{S})\right\rangle=c_{l_1} c_{l_2}\sum_{L=0}^{\infty}\sum_{M=-L}^{L}g_{L,M} C^{L,l_1,l_2}_{M,m_1,m_2}
\end{align}
with
\begin{align}
    C^{L,l_1,l_2}_{M,m_1,m_2}&=\int_{-1}^1d \cos x\int_0^{2\pi}d y~Y_{L,M}(\hat{S})~Y_{l_1,m_1}(\hat{S})~Y_{l_2,m_2}(\hat{S}) \\
    &=\sqrt{\frac{(2L+1)(2l_1+1)(2l_2+1)}{4\pi}}
\begin{pmatrix}
L & l_1 & l_2 \\
0 & 0 & 0
\end{pmatrix}
\begin{pmatrix}
L & l_1 & l_2 \\
M & m_1 & m_2
\end{pmatrix}~.
\end{align}
The coefficient $C^{L,l_1,l_2}_{M,m_1,m_2}$ is nonzero only if the following conditions are simultaneously satisfied:
\begin{itemize}
    \item $|l_1-l_2|\le L\le l_1+l_2$.
    \item $L+l_1+l_2$ is even.
    \item $M+m_1+m_2=0$.
\end{itemize}
As a consequence,
\begin{align}
\mathcal{C}^\prime_{l_1,m_1,l_2,m_2}=\mathcal{C}^\prime_{l_2,m_2,l_1,m_1}~.
\end{align}

Table~\ref{tab:1} also summarizes the dependence of $\mathcal{C}^\prime_{l_1,m_1,l_2,m_2}$ on the coefficients $c_i$ and $g_{L,M}$ for those potentially non-vanishing $\mathcal{C}_{l_1,m_1,l_2,m_2}$ with $l_1\le l_2$. Since $\mathcal{C}_{l_1,m_1,l_2,m_2}=0$ whenever $m_1+m_2\ne 0$, the corresponding nonzero expressions for $\mathcal{C}^\prime_{l_1,m_1,l_2,m_2}$ do not depend on coefficients $g_{L,M}$ with $M\ne 0$.

\subsection{Existence conditions for LHVT constructions}\label{sec:3-2}

I now determine the region of the parameter space $(a_1,a_2,a_3,b_2,b_3)$ for which $\mathcal{W}^\text{LHVT}_s(\theta_1,\phi_1,\theta_2,\phi_2)$ can reproduce $\mathcal{W}^\text{QFT}_s(\theta_1,\phi_1,\theta_2,\phi_2)$. To keep the discussion as general as possible, I allow $\theta_W$ to vary freely over the interval $(0,\frac{\pi}{2})$.

Before turning to the generic case, let us first note a trivial class of configurations for which an LHVT can always be constructed. If one of the three $a_i~(i=1,2,3)$ is equal to $\pm1$ while the other two vanish, then the $Z_1Z_2$ state is a product state, and an explicit LHVT realization is immediate. For the case $a_k=\pm1$, one convenient choice is
\begin{align}
& G(x,y)=\frac{1}{2\pi}\delta(\cos x-1)~, \quad F(\cos\theta)=\frac{3}{8\pi}\sum_{\lambda=\pm 1} (1-\lambda w)d^{1}_{\lambda,k-2}(\theta)d^{1}_{\lambda,k-2}(\theta)
~.
\end{align}
In what follows, I therefore restrict attention to the nontrivial situation in which at most one of the three $a_i~(i=1,2,3)$ vanishes.

\subsubsection{The case $w\ne 0$}\label{sec:3-2-1}

A necessary condition for an LHVT construction is that the QFT coefficients satisfy
\begin{align}
&\mathcal{C}_{l_1,m_1,l_2,m_2}=\mathcal{C}_{l_2,m_2,l_1,m_1}=\mathcal{C}_{l_1,m_1,l_2,m_2}^*~. \label{ccstar}
\end{align}
For $w\ne 0$, Table~\ref{tab:1} then implies
\begin{align}
& a_1a_2\sin b_2=0~,\quad a_1a_3\sin b_3=0~,\quad a_2a_3\sin(b_2-b_3)=0~.
\end{align}
The possible cases can thus be classified as follows
\begin{itemize}
    \item If all three $a_i~(i=1,2,3)$ are nonzero, one must have $b_2=b_3=0~$.
    \item If one of the three $a_i~(i=1,2,3)$ vanishes, one still finds that $b_2=b_3=0$ can be chosen without loss of generality. For example, when $a_1=0$, the condition $a_2a_3\sin(b_2-b_3)=0$ requires $b_2=b_3$, and one can then set $b_2=b_3=0$ without loss of generality.
\end{itemize}

I therefore set $b_2=b_3=0$ in the following. Equating $\mathcal{C}_{l_1,m_1,l_2,m_2}$ and $\mathcal{C}^{\prime}_{l_1,m_1,l_2,m_2}$ in Table~\ref{tab:1} leads to the system
\begin{align}
    & c_1 g_{1,0}=\frac{3}{4} \sqrt{\frac{3}{\pi }} w \left(a_1^2-a_3^2\right)~, \\
&
c_2 g_{2,0}=\frac{1}{4} \sqrt{\frac{5}{\pi }} \left(1-3 a_2^2\right)~,\\
&
c_1^2=\frac{9}{4} w^2 \left(1-a_2(2a_1+a_2+2a_3)\right)~,\\
&
c_1^2 g_{2,0}=\frac{9}{8} \sqrt{\frac{5}{\pi }} w^2 \left(1+a_2(a_1-a_2+a_3)\right)~,\\
&
c_1 c_2 g_{1,0}=\frac{3}{8} \sqrt{\frac{3}{\pi }} w (a_1-a_3) (a_1-3 a_2+a_3)~,\\
&
c_1 c_2 g_{3,0}=\frac{3}{8} \sqrt{\frac{7}{\pi }} w (a_1-a_3) (a_1+2 a_2+a_3)~,\\
&
c_2^2=\frac{1}{4} \left(1-3a_2(2a_1-a_2+2a_3)+12 a_1 a_3\right)~,\\
&
c_2^2 g_{4,0}=\frac{3 (a_1+2 a_2+a_3)^2}{8 \sqrt{\pi }}~,\\
&
c_2^2 g_{2,0}=\frac{1}{8} \sqrt{\frac{5}{\pi }} \left(1-3a_2(a_1-a_2+a_3)-12a_1a_3\right)~.
\end{align}
Imposing these relations, together with the additional requirement that all $\mathcal{C}^{\prime}_{l_1,m_1,l_2,m_2}$ with $l_1\le l_2$ that are not listed in Table~\ref{tab:1} vanish, yields a unique solution,
\begin{align}
   & a_1=a_3=\frac{1}{\sqrt{3}}~,\quad a_2=-\frac{1}{\sqrt{3}}~,\label{jct1}\\
   & c_1=\pm \frac{3}{\sqrt{2}}w~,\quad c_2=\sqrt{\frac{5}{2}}~,\quad c_{l>2}=0~,\\
   & g_{1,0}=g_{2,0}=g_{3,0}=g_{4,0}=0~. \label{g1234}
\end{align}
The $Z_1Z_2$ state given in Eq.~(\ref{jct1}) is entangled, which leads to $8\langle\cos(2\phi_1- 2\phi_2)\rangle=\frac{2}{3}$.
The condition $c_{l>2}=0$ is in fact necessary, since any non-vanishing $c_l$ with $l>2$ would inevitably generate nonzero  $\mathcal{C}^{\prime}_{0,0,l,0}$, $\mathcal{C}^{\prime}_{1,m,l,-m}$,
$\mathcal{C}^{\prime}_{2,m,l,-m}$, or
$\mathcal{C}^{\prime}_{l,m,l,-m}$.

The simplest choice of $G(x,y)$ consistent with Eq.~(\ref{g1234}) is
\begin{align}
    G(x,y)=\frac{1}{4\pi}~.
\end{align}
The corresponding decay distribution is then
\begin{align}
    F(\cos\theta)=\frac{1}{4\pi}\left(1+c_1 P_1(\cos\theta)+c_2 P_2(\cos\theta)\right)~.
\end{align}
Requiring $F(\cos\theta)\ge 0$ gives
\begin{align}
   w\in [w_-,w_+]~, \quad i.e.~ \theta_W\in [v,u_{-}]\cup [u_{+},\frac{\pi}{2})~, \label{wwind}
\end{align}
where
\begin{align}
    & w_{\pm}=\pm \sqrt{\frac{2\sqrt{10}-5}{3}}~,\\
    & \frac{-1+2\cos(2{v})}{2-2\cos(2v)+\cos(4v)}=w_+~,\\
    & \frac{-1+2\cos(2{u_{\pm}})}{2-2\cos(2u_{\pm})+\cos(4u_{\pm})}=w_-~,\quad u_{-}<u_{+}~.
\end{align}
Numerically, one finds
\begin{align}
    & v\approx 23.18^{\circ}~,\quad u_{-}\approx 35.97^{\circ}~, \quad u_{+}\approx 72.28^{\circ}~.
\end{align}

According to the PDG \cite{ParticleDataGroup:2024cfk}, for leptonic $Z$ decays, the effective leptonic weak mixing angle is measured to be $\sin^2\theta_{\rm eff}^{\ell}=0.23148\pm0.00012$, corresponding to $\theta_{\rm eff}^{\ell}=28.759^\circ\pm0.008^\circ$. Therefore, the measured value of $\theta_{\rm eff}^{\ell}$ lies within the windows specified by Eq.~(\ref{wwind}) and is very close to $\frac{1}{2}(v+u_{-})\approx 29.58^{\circ}$.

\subsubsection{The case $w= 0$}\label{sec:3-2-2}

The condition $w=0$ implies
\begin{align}
   \theta_W=30^{\circ}~.
\end{align}
In this case, Eq.~(\ref{ccstar}) reduces to
\begin{align}
& a_1 a_3 \sin b_3=0~,\quad a_2 a_1 \sin b_2-a_2 a_3 \sin (b_2-b_3)=0~.
\end{align}
Without loss of generality, only the following two cases need to be considered:
\begin{itemize}
    \item $b_2=b_3=0$. The matching conditions become 
    \begin{align}
    & c_1=0~,\quad c_{l>2}=0~,\label{case1-1}\\
    &
c_2 g_{2,0}=\frac{1}{4} \sqrt{\frac{5}{\pi }} \left(1-3 a_2^2\right)~,\label{case1-2}\\
&
c_2^2=\frac{1}{4} \left(1-3a_2(2a_1-a_2+2a_3)+12 a_1 a_3\right)~,\label{case1-3}\\
&
c_2^2 g_{4,0}=\frac{3 (a_1+2 a_2+a_3)^2}{8 \sqrt{\pi }}~,\label{case1-4}\\
&
c_2^2 g_{2,0}=\frac{1}{8} \sqrt{\frac{5}{\pi }} \left(1-3a_2(a_1-a_2+a_3)-12a_1a_3\right)~.\label{case1-5}
\end{align}
\item $b_3=0$ and $a_1=a_3=\sqrt{\frac{1-a_2^2}{2}}$. The corresponding system is
\begin{align}
    & c_1=0~,\quad c_{l>2}=0~,\label{case2-1}\\
    &
c_2 g_{2,0}=\frac{1}{4} \sqrt{\frac{5}{\pi }} \left(1-3 a_2^2\right)~, \label{case2-2}\\
&
c_2^2=\frac{7}{4}-\frac{3}{4}a_2^2-\frac{3}{\sqrt{2}}a_2\sqrt{1-a_2^2}\cos b_2~,\label{case2-3}\\
&
c_2^2 g_{4,0}=\frac{3}{4\sqrt{\pi}}\left(1+a_2^2+2\sqrt{2}a_2\sqrt{1-a_2^2}\cos b_2\right)~,\label{case2-4}\\
&
c_2^2 g_{2,0}=-\frac{1}{8} \sqrt{\frac{5}{\pi }} \left(5-9a_2^2+3\sqrt{2}a_2\sqrt{1-a_2^2}\cos b_2\right)~. \label{case2-5}
\end{align}
\end{itemize}
In either case, the requirement $c_{l>2}=0$ is actually indispensable, because any nonzero $c_l$ with $l>2$ would necessarily lead to non-vanishing $\mathcal{C}^{\prime}_{0,0,l,0}$, $\mathcal{C}^{\prime}_{2,m,l,-m}$, or $\mathcal{C}^{\prime}_{l,m,l,-m}$.

For both cases, determining the exact parameter region that admits an LHVT construction requires analyzing rather complicated functions and the locations of their roots. I therefore content ourselves with a criterion that is both necessary and sufficient for the existence of an LHVT. For either case, I denote by $\chi_1~,\chi_2~,\chi_3~,\chi_4$ the right-hand sides of Eqs.~(\ref{case1-2}), (\ref{case1-3}), (\ref{case1-4}), and (\ref{case1-5}), or equivalently those of Eqs.~(\ref{case2-2}), (\ref{case2-3}), (\ref{case2-4}), and (\ref{case2-5}), respectively.

For the system to admit a solution, one must first require
\begin{align}
    & \chi_4^2-\chi_1^2\chi_2=0~.\label{pj1}
\end{align}
A direct calculation shows that $\chi_2=0$ leads to no solution in either case, so one must also impose
\begin{align}    
& \chi_2>0~. \label{pj2}
\end{align}
Moreover, $\chi_4=\chi_1=0$ yields, in both cases, only the solution $a_1=a_3=-a_2=\frac{1}{\sqrt{3}}$ and $b_2=b_3=0$, which has already been discussed above ($0\in [w_-,w_+]$). To avoid repeating that case, I further require
\begin{align}
    & \chi_{1}\chi_{4}\ne 0~. \label{pj3}
\end{align}

When Eqs.~(\ref{pj1})--(\ref{pj3}) are satisfied simultaneously, one obtains
\begin{align}    
& c_2=\frac{\chi_1 \chi_2}{\chi_4}~,\quad  g_{2,0}=\frac{\chi_4}{\chi_2}~,\quad g_{4,0}=\frac{\chi_3}{\chi_2}~.
\end{align}
To ensure the non-negativity of
\begin{align}
    F(\cos\theta)=\frac{1}{4\pi}\left(1+c_2 P_l(\cos\theta)\right)~,
\end{align}
one must further impose
\begin{align}
    & c_2\in [-1,2]~. \label{pj4}
\end{align}

\begin{figure}[tbp]
\begin{center}
\includegraphics[width=0.49\textwidth]{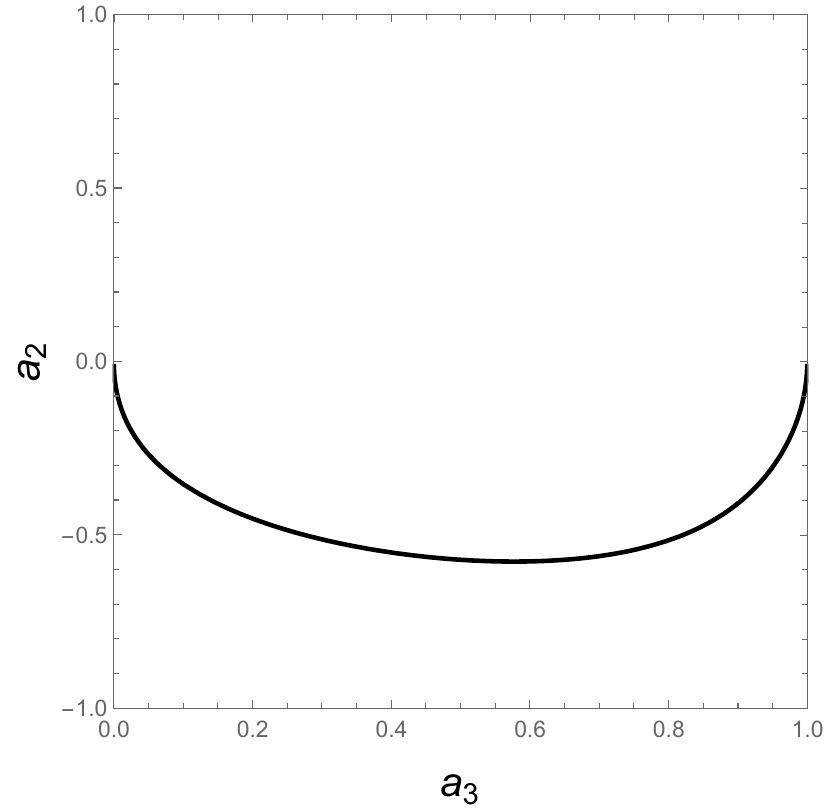}~
\includegraphics[width=0.49\textwidth]{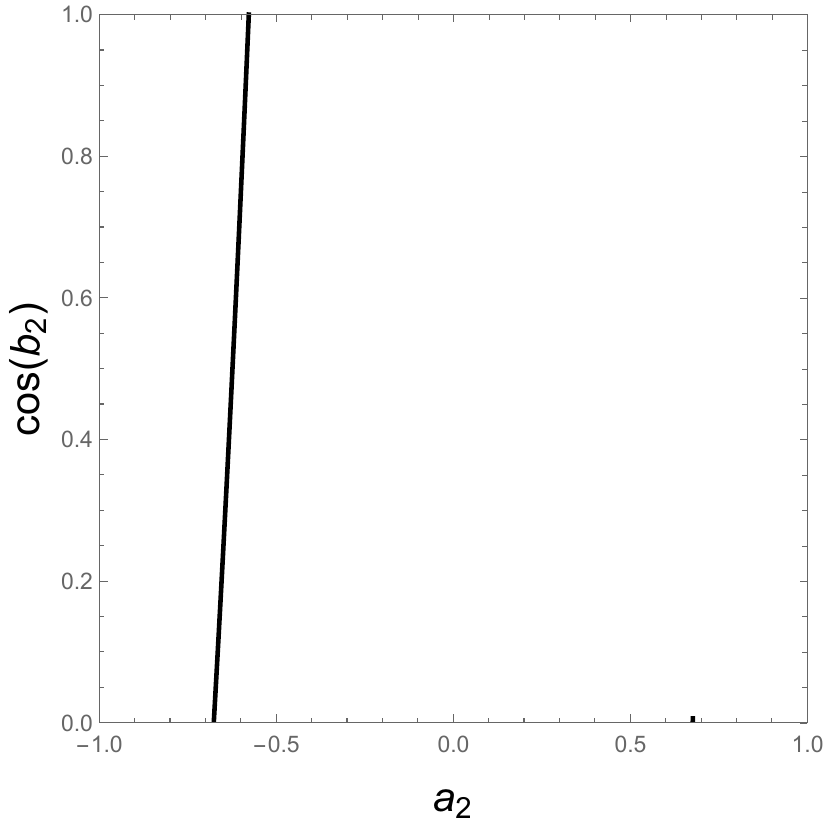}
\end{center}
 \vspace*{-0.1in}
\caption{Black curves denote the allowed parameter space for constructing LHVT at $w=0$, without including the points corresponding to product states. Left: the case $b_2=b_3=0$ with $a_1=\sqrt{1-a_2^2-a_3^2}$. Right: the case $b_3=0$ with $a_1=a_3=\sqrt{\frac{1-a_2^2}{2}}$.
The black curves in both the left and right panels include the point defined by $a_1=a_3=-a_2=\frac{1}{\sqrt{3}}$ and $b_2=b_3=0$.
}\label{a2a3b2}
\end{figure}

The remaining conditions arise from the existence of a nonnegative distribution $G(x,y)$. Since
\begin{align}
    & g_{2,0}=\frac{1}{4}\sqrt{\frac{5}{\pi}}(-1+3\langle\cos^2 x\rangle)~, \\
    & g_{4,0}=\frac{3}{16\sqrt{\pi}}(3-30\langle\cos^2 x\rangle+35\langle\cos^4 x\rangle)~,
\end{align}
I obtain
\begin{align}
    & \langle\cos^2 x\rangle=t_1=\frac{1}{15}(5+4g_{2,0}\sqrt{5\pi})~, \\
    & \langle\cos^4 x\rangle=t_2=\frac{1}{105}(21+24g_{2,0}\sqrt{5\pi}+16g_{4,0}\sqrt{\pi})~.
\end{align}
Clearly,
\begin{align}
    & t_1\in [0,1]~. \label{pj5}
\end{align}
must hold. In addition, the Cauchy–Schwarz inequality gives
\begin{align}
    \langle\cos^2 x\rangle^2\le\langle\cos^4 x\rangle\le \langle\cos^2 x\rangle~,
\end{align}
so one must also require
\begin{align}
    & t_2\in [t_1^2,t_1]~. \label{pj6}
\end{align}
When Eqs.~(\ref{pj5}) and (\ref{pj6}) are satisfied, an explicit construction of $G(x,y)$ can be given. For $t_1\in [0,1)$, one may choose
    \begin{align}
   & G(x,y)=\frac{1}{2\pi}\left(\rho\delta(\cos x-1)+(1-\rho)\delta (\cos x-\sqrt{t_1^\prime})\right)
\end{align}
with
\begin{align}
   & \rho=\frac{t_2-t_1^2}{1+t_2-2t_1}~, \quad  t_1^\prime= \frac{t_1-t_2}{1-t_1}~.
\end{align}
For $t_1=t_2=1$, one may take
\begin{align}
   & G(x,y)=\frac{1}{2\pi}\delta(\cos x-1)~.
\end{align}
Therefore, when $w=0$ and $a_2\ne -\frac{1}{\sqrt{3}}$, Eqs.~(\ref{pj1}), (\ref{pj2}), (\ref{pj3}), (\ref{pj4}), (\ref{pj5}), and (\ref{pj6}) provide a necessary and sufficient criterion for the existence of an LHVT construction. In Figure~\ref{a2a3b2}, the black curves represent the allowed parameter space satisfying this criterion at $w=0$, excluding the points corresponding to product states. In both the left and right panels, I have included the point defined by $a_1=a_3=-a_2=\frac{1}{\sqrt{3}}$ and $b_2=b_3=0$ on the black curves. 
In the left panel, the points on the black curves correspond to $a_3\in(0,1)$, with $a_2$ and $a_3$ satisfying Eq.~(\ref{pj1}).
In the right panel, the isolated point corresponds to $a_2=\sqrt{\frac{\sqrt{33}-3}{6}}$ and $b_2=\frac{\pi}{2}$, while the continuous curve is given by
\begin{align}
    \cos{b_2}=\frac{\left(3 a_2^2-2\right) \sqrt{1-a_2^2}+\left(3 a_2^2-1\right) \sqrt{2-a_2^2}}{\sqrt{2} a_2}~, \quad -\sqrt{\frac{\sqrt{33}-3}{6}}\le a_2\le -\frac{1}{\sqrt{3}}~.
\end{align}
Since the points corresponding to product states are not included on the black curves in Figure~\ref{a2a3b2}, every point on these curves corresponds to an entangled state.

Finally, let us consider the phenomenologically relevant case in which the interaction between $s$ and $Z$ is of the form
\begin{align}
    s Z^\mu Z_\mu~.
\end{align}
A straightforward calculation based on QFT shows that the $Z_1 Z_2$ state produced in the process $s\to Z_1 Z_2$ is entangled and characterized by
\begin{align}
    a_1=a_3=\frac{1}{\sqrt{2+\sigma^2}}~,\quad a_2=\frac{\sigma}{\sqrt{2+\sigma^2}}~,\quad b_2=b_3=0
\end{align}
with
\begin{align}
    \sigma=1-\frac{m_s^2}{2 m_Z^2}~.
\end{align}
In this case, if $m_s=2m_Z$, an LHVT construction exists regardless of whether $w=0$ or not. By contrast, if $m_s>2m_Z$, no LHVT construction exists for either $w=0$ or $w\ne 0$; in particular, when $w=0$, the above criteria cannot be satisfied simultaneously, as can also be seen from Figure~\ref{a2a3b2}.

\section{Conclusion} \label{sec:con} 

In this work, I have investigated whether the angular correlations in the leptonic decays of $Z_1(\to e^-_1 e^+_1)Z_2(\to e^-_2 e^+_2)$, where the $Z_1Z_2$ pair is produced in the decay of a spin-0 particle, can be reproduced by an LHVT under angular-momentum conservation. Our analysis was carried out by formulating both the QFT prediction and the LHVT description of the joint angular distribution of $e^-_1$ and $e^-_2$ in a common spherical-harmonic basis, which makes it possible to compare the two frameworks coefficient by coefficient at the level of the fully differential leptonic angular distribution.

Within QFT, angular-momentum conservation restricts the polarization state of the pure $Z_1Z_2$ system to a superposition of the helicity states $|-1,-1\rangle$, $|0,0\rangle$, and $|1,1\rangle$. On the LHVT side, I constructed the most general hidden-variable description compatible with angular-momentum conservation by introducing a distribution $G(x,y)$ for the hidden spin direction and a decay response function $F(\hat{S}_i\cdot\hat{e}_i)$ for each vector boson. Expanding these functions in spherical harmonics and Legendre polynomials, respectively, allowed us to derive explicit matching conditions between the QFT coefficients $\mathcal{C}_{l_1,m_1,l_2,m_2}$ and their LHVT counterparts $\mathcal{C}^\prime_{l_1,m_1,l_2,m_2}$.

Our results show that the existence of an LHVT construction is highly constrained. For $w\neq 0$, apart from the trivial product-state cases, the matching conditions admit only a unique entangled solution, $a_1=a_3=-a_2=\frac{1}{\sqrt{3}}$ and $b_2=b_3=0$, together with a restricted interval of the weak-mixing angle $\theta_W$. For $w=0$, I derive a complete criterion for LHVT existence in terms of a set of algebraic consistency conditions and positivity constraints on both the decay response function and the hidden-variable distribution. This criterion is necessary and sufficient, and I also provide an explicit construction of $G(x,y)$ whenever the conditions are satisfied.

As an illustrative application, I study the case in which the parent spin-0 particle couples to the $Z$ bosons through the interaction $sZ^\mu Z_\mu$. The resulting $Z_1Z_2$ polarization state depends only on the mass ratio $m_s/m_Z$. I find that an LHVT construction exists at threshold, $m_s=2m_Z$, but fails for $m_s>2m_Z$, irrespective of whether $w=0$ or $w\neq 0$.

More broadly, the framework developed here provides a systematic method for testing the classical simulability of spin-correlation observables in unstable di-boson systems. It would be interesting to extend this analysis to mixed $Z_1Z_2$ states, to other production mechanisms, and to more general two-vector systems such as $W^+W^-$ or $Z\gamma$, where different spin structures and decay patterns may lead to qualitatively new possibilities for entanglement and for its distinction from local hidden-variable models.

\begin{acknowledgments}

J. Pei is supported by the National Natural Science Foundation of China (Project No. 12505121), by the Joint Fund of Henan Province Science and Technology R$\&$D Program (Project No. 245200810077), by the Startup Research Fund of Henan Academy of Sciences (Project No. 20251820001), and by the Scientific and Technological Research Project of Henan Academy of Sciences (Project No. 20262320001).

\end{acknowledgments}

\bibliographystyle{jhep}
\bibliography{jhep}

\end{document}